\documentclass[twocolumn]{aastex63}

\received{September 24, 2019}
\revised{August 10, 2020}
\accepted{\today}

\shorttitle{Non-Newtonian Gravity in Strange Quark Stars}
\shortauthors{Yang et al.}

\begin{document}

\title{Non-Newtonian Gravity in Strange Quark Stars and Constraints
  from the Observations of PSR J0740+6620 and GW170817}

\correspondingauthor{Shu-Hua Yang}
\email{ysh@mail.ccnu.edu.cn}

\author{Shu-Hua Yang}
\affiliation{Institute of Astrophysics, Central China Normal
	University, Wuhan 430079, China}

\author{Chun-Mei Pi}
\affiliation{School of Physics and Mechanical \& Electrical
	Engineering, Hubei University of Education, Wuhan 430205, China}

\author{Xiao-Ping Zheng} \affiliation{Institute of Astrophysics,
  Central China Normal University, Wuhan 430079, China}
\affiliation{Department of Astronomy, Huazhong University of Science
  and Technology, Wuhan 430074, China}

\author{Fridolin Weber} \affiliation{Department of Physics, San Diego
  State University, San Diego, CA 92182, United States of America}
\affiliation{Center for Astrophysics and Space Sciences, University of
  California at San Diego, La Jolla, CA 92093, United States of
  America}

\begin{abstract}
	We investigate the effects of non-Newtonian gravity on the
        properties of strange quark stars (QSs) and constrain the
        parameters of the standard MIT bag model used to
          describe strange quark matter (SQM) by employing the mass of
          PSR J0740+6620 and the tidal deformability of GW170817. We
          find that, for the standard MIT bag model, these mass and
          tidal deformability observations would rule out the
          existence of QSs if non-Newtonian gravity effects are
          ignored. For a strange quark mass of $m_{s}=95$ MeV, we
        find that QSs can exist for values of the non-Newtonian
        gravity parameter $g^{2}/\mu^{2}$ in the range of 1.37
        GeV$^{-2}\leq g^{2}/\mu^{2}\leq$ 7.28 GeV$^{-2}$ and limits
        on the bag constant and the strong interaction coupling 
        constant of the SQM model given by 141.3 MeV$\leq
        B^{1/4}\leq$ 150.9 MeV and $\alpha_{S}\leq 0.56$.  For a
        strange quark mass of $m_{s}=150$ MeV, QSs can exist for 1.88
        GeV$^{-2}\leq g^{2}/\mu^{2}\leq$ 6.27 GeV$^{-2}$ and limits
        on the parameters of the SQM model given by 139.7 MeV$\leq
        B^{1/4}\leq$ 147.3 MeV and $\alpha_{S}\leq 0.49$.
\end{abstract}



\section{Introduction}

The gravitational wave event GW170817 \citep{2017PhRvL.119p1101A} and
its associated electromagnetic counterpart \citep{2017ApJ...848L..12A}
have placed constraints on the neutron star matter equation of state
(EOS) \citep[see reviews,
e.g.,][]{2019JPhG...46g3002O,2019EPJA...55...80R,2019EPJA...55..117L,2019PrPNP.10903714B}.
Among these studies, several works have checked whether the
observations from GW170817 are compatible with QSs, in which the
matter consists of deconfined up ($u$), down ($d$), strange ($s$)
quarks and electrons
\citep{2018PhRvD..97h3015Z,2018RAA....18...24L,2019EPJA...55...60L}.

\cite{2018PhRvD..97h3015Z} found that the tidal deformability of
GW170817, together with the mass of PSR J0348+0432 ($2.01 \pm 0.04\,
M_{\odot}$) \citep{2013Sci...340..448A}, can restrict the parameter
space of the SQM model, but can not rule out the possible
existence of QSs.  However, the dimensionless tidal deformability for
a $1.4\, M_{\odot}$ star ($\Lambda(1.4)$) employed by these authors is
$\Lambda(1.4)\leq800$ \citep{2017PhRvL.119p1101A}, which has been
improved to $\Lambda(1.4)=190_{-120}^{+390}$ ($\Lambda(1.4)\leq 580$
will be used in this paper) \citep{2018PhRvL.121p1101A}.  Moreover,
the millisecond pulsar J0740+6620 with a mass of
$2.14_{-0.09}^{+0.10}\, M_{\odot}$ (68.3\% credibility interval;
$2.14_{-0.18}^{+0.20}\, M_{\odot}$ for a 95.4\% credibility
interval) was reported recently \citep{2020NatAs...4...72C}, which
sets a new record for the maximum mass of neutron stars (NSs). In
this paper, we show that the existence of QSs seems to be ruled out by
these new data (see panel (a) in Figs.\ \ref{fig2} and \ref{fig3}.)
if the standard MIT bag model of SQM is used to compute
the bulk properties of SQs.

This is no longer the case, however if
non-Newtonian gravity is considered, as will be shown in this paper.
Effects of non-Newtonian gravity on the properties of NSs and QSs
have been studied
\citep[e.g.,][]{2009PhRvD..79l5023K,2009PhRvL.103u1102W,2011MPLA...26..367S,2014JPhG...41g5203L,2017RAA....17...11L},
and it was found that the inclusion of non-Newtonian gravity leads
to stiffer EOSs and higher maximum masses of compact stars. Hence,
within the framework of non-Newtonian gravity, the observed massive
pulsars do not rule out a rather soft behavior of the EOS of dense
nuclear matter \citep{2009PhRvL.103u1102W}.

The conventional inverse-square-law of gravity is
expected to be violated in the efforts of trying to unify gravity
with the other three fundamental forces, namely, the
electromagnetic, weak and strong interactions
\citep{1999snng.book.....F,2003ARNPS..53...77A,2009PrPNP..62..102A}. Non-Newtonian
gravity arise due to either the geometrical effect of the extra
space-time dimensions predicted by string theory and/or the exchange
of weakly interacting bosons, such as a neutral very weakly coupled
spin-1 gauge U-boson proposed in the super-symmetric extension of
the standard model
\citep{1980PhLB...95..285F,1981NuPhB.187..184F}. Non-Newtonian
gravity is often characterized effectively by adding a Yukawa term
to the normal gravitational potential \citep{1971NPhS..234....5F}.
Constraints on the deviations from Newton's gravity have been set
experimentally, see \cite{2015CQGra..32c3001M} and references
therein. (Besides, an extra Yukawa term also naturally arises in the
weak-field limit of some modified theories of gravity, e.g., f(R) gravity,
the nonsymmetric gravitational theory, and
Modified Gravity \citep{2014JPhG...41g5203L,2019EPJA...55..117L}.)

In this paper, we will study the effects of non-Newtonian gravity on
the properties of QSs and constrain the parameter space of the
SQM model using the updated tidal deformability of GW170817 and
the recently reported mass of PSR J0740+6620.

This paper is organized as follows: In Sec.\ \ref{SecII}, we briefly
review the theoretical frame work of the EOS of strange quark matter
including the non-Newtonian gravity effects and the calculations of
the structure and tidal deformability of strange stars. In Sec.\ \ref{ResDisc},
numerical results and discussions are presented. Finally, a summary is
given in Sec.\ \ref{Summary}.

\section{Theoretical Framework}\label{SecII}

\subsection{EOS of Strange Quark Matter Including the Non-Newtonian
	Gravity Effects}

Before discussing the effects of non-Newtonian gravity on the EOS of
SQM
\citep{1984PhRvD..30.2379F,1986A&A...160..121H,1986ApJ...310..261A,alcock88,madsen99},
we briefly review the phenomenological model for the EOS employed in
this paper, namely, the standard bag model
\citep{1984PhRvD..30.2379F,1986A&A...160..121H,1986ApJ...310..261A,2005PrPNP..54..193W}. In
that model, $u$ and $d$ quarks are treated as massless particles but
$s$ quarks have a finite mass, $m_s$.  First-order perturbative
corrections in the strong interaction coupling constant $\alpha_{S}$
are taken into account.

The thermodynamic potential for the $u$, $d$ and $s$ quarks, and for
the electrons are given by
\citep{1986ApJ...310..261A,2015RAA....15..871P}
\begin{equation}
\Omega_{u}=-\frac{\mu_{u}^{4}}{4\pi^{2}}\bigg(1-\frac{2\alpha_{S}}{\pi}\bigg),
\end{equation}

\begin{equation}
\Omega_{d}=-\frac{\mu_{d}^{4}}{4\pi^{2}}\bigg(1-\frac{2\alpha_{S}}{\pi}\bigg),
\end{equation}

\begin{eqnarray}
\nonumber
\Omega_{s}&=&-\frac{1}{4\pi^{2}}\bigg\{\mu_{s}\sqrt{\mu_{s}^{2}-m_{s}^{2}}
(\mu_{s}^{2}-\frac{5}{2}m_{s}^{2})+\frac{3}{2}m_{s}^{4}f(u_{s},m_{s})\\ \nonumber
&&-\frac{2\alpha_{S}}{\pi}\bigg[3\bigg(\mu_{s}\sqrt{\mu_{s}^{2}-m_{s}^{2}}
-m_{s}^{2}f(u_{s},m_{s})\bigg)^{2}\\ \nonumber
&&-2(\mu_{s}^{2}-m_{s}^{2})^{2}-3m_{s}^{4}\textrm{ln}^{2}\frac{m_{s}}{\mu_{s}}\\
&&+6\textrm{ln}\frac{\sigma}{\mu_{s}}\bigg(\mu_{s}m_{s}^{2}\sqrt{\mu_{s}^{2}
	-m_{s}^{2}}-m_{s}^{4}f(u_{s},m_{s})\bigg)\bigg]\bigg\},
\label{OmegaS}
\end{eqnarray}

\begin{equation}
\Omega_{e}=-\frac{\mu_{e}^{4}}{12\pi^{2}},
\end{equation}
where
$f(u_{s},m_{s})\equiv\textrm{ln}[(\mu_{s}+\sqrt{\mu_{s}^{2}-m_{s}^{2}})/m_{s}]$. The
quantity $\sigma$ (= 300 MeV) is a renormalization constant whose
value is of the order of the chemical potential of strange quarks,
$\mu_s$.  Values of $m_s=95$ MeV and $m_s=150$ MeV have been
considered for the strange quark mass in our calculations
\citep{2014ChPhC..38i0001O,Tanabashi18}.

The number density of each quark species is given by
\begin{equation}
n_{i}=-\frac{\partial\Omega_{i}}{\partial\mu_{i}},
\end{equation}
where $\mu_{i}$ ($i=u,d,s,e$) are the chemical potentials. For SQM,
chemical equilibrium is maintained by the
weak-interaction, which leads for the chemical potentials to the
following conditions,
\begin{eqnarray}
\mu_{d} &=&  \mu_{s} , \\
\mu_{s} &=& \mu_{u}+\mu_{e} .
\end{eqnarray}
The electric charge neutrality condition is given by
\begin{equation}
\frac{2}{3}n_{u}-\frac{1}{3}n_{d}-\frac{1}{3}n_{s}-n_{e}=0.
\end{equation}
The total baryon number density follows from
\begin{equation}
n_{b}=\frac{1}{3}(n_{u}+n_{s}+n_{d}) .
\end{equation}
The energy density without the effects of the non-Newtonian gravity is given by
\begin{equation}
\epsilon_{Q}=\sum_{i=u,d,s,e}(\Omega_{i}+\mu_{i}n_{i})+B,
\label{eq:epsQ}
\end{equation}
and the corresponding pressure is obtained from
\begin{equation}
p_{Q}=-\sum_{i=u,d,s,e}\Omega_{i}-B ,
\label{eq:pQ}
\end{equation}
where $B$ denotes the bag constant.

According to \cite{1971NPhS..234....5F}, non-Newtonian
gravity can be described by adding a Yukawa-like term to the conventional
gravitational potential between two objects with masses $m_{1}$ and $m_{2}$, i.e.,
\begin{equation}
V(r)=-\frac{G_{\infty}m_{1}m_{2}}{r} \left( 1+\alpha e^{-r/\lambda}
\right) = V_N(r) + V_Y(r),
\label{vr}
\end{equation}
where $V_Y(r)$ is the Yukawa correction to the Newtonian potential
$V_N(r)$.  The quantity $G_{\infty} = 6.6710\times 10^{-11}~\rm{N} \,
{\rm m}^2/{\rm kg}^2$ is the universal gravitational constant,
$\alpha$ is the dimensionless coupling constant of the Yukawa force,
and $\lambda$ is the range of the Yukawa force mediated by the
exchange of bosons of mass $\mu$ (given in natural units) among $m_1$
and $m_2$,
\begin{equation}
\lambda=\frac{1}{\mu}.
\label{mulam}
\end{equation}
In this picture, the Yukawa term is the static limit of an interaction
mediated by virtual bosons.  The strength parameter in Eq.\ (\ref{vr})
is given by
\begin{equation}
\alpha=\pm \frac{g^{2}}{4\pi G_{\infty}m_{b}^{2}},
\end{equation}
where the $\pm$ sign refers to scalar (upper sign) or vector
(lower sign) bosons, $g$ is the boson-baryon coupling constant, and
$m_{b}$ is the baryon mass.

\begin{figure}[htb]
\begin{center}
\includegraphics[width=0.55\textwidth]{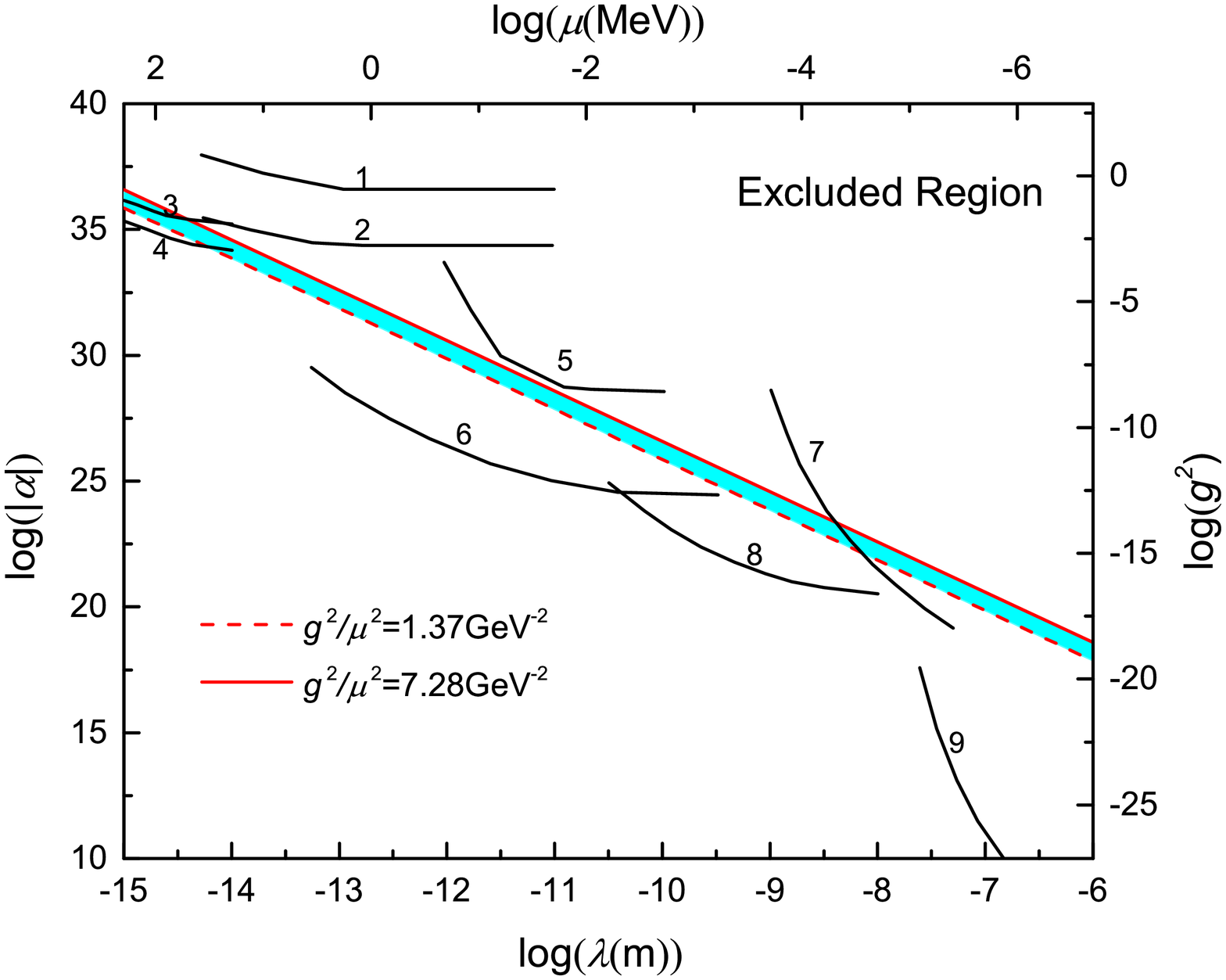}
\caption{Upper bounds on the strength parameter
    $|\alpha|$ respectively the boson-nucleon coupling constant $g$ as
    a function of the range of the Yukawa force $\mu$ (bottom) and the
    mass of hypothetical bosons (top), set by different experiments:
    curves 1 and 2 refer to constraints from np scattering of scalar and
    vector bosons, respectively \citep{Kamyshkov2008}; 3 and 4 are
    constraints extracted from charge radii and binding energies of atomic
    nuclei, respectively \citep{Xu2013}; 5 was established from the
    spectroscopy of antiprotonic He atoms and 6 from neutron total
    cross section scattering of $^{208}{\rm Pb}$ nuclei
    \citep{Pokotilovski2006}; 7 is from an experiment measuring the
    Casimir force between a Au-coated microsphere and a silicon
    carbide plate \citep{Klimchitskaya2020}; 8 is obtained by
    measuring the angular distribution of 5~\AA ~neutrons scattered
    off of an atomic xenon gas \citep{Kamiya2015}; 9 shows the
    constraints from the force measurements between a test mass and
    rotating source masses of gold and silicon \citep{Chen2016}. The
    cyon-shaded area corresponds to 1.37 GeV$^{-2}\leq
    g^{2}/\mu^{2}\leq$ 7.28 GeV$^{-2}$. } \label{fig0}
	\end{center}
\end{figure}

  Shown in Fig.\ \ref{fig0} are the parameter spaces $\log
  g^{2}$--$\log\mu$ and $\log|\alpha|$--$\log\lambda$ associated with
  the above mentioned hypothetical bosons. Constraints on the
  parameter spaces set by several recent experiments are indicated.
  The theoretical bounds on $g^{2}/\mu^{2}$ of 1.37 GeV$^{-2}\leq
  g^{2}/\mu^{2}\leq$ 7.28 GeV$^{-2}$, for which QSs are found to exist
  (see Sect.~\ref{ResDisc}) are shown by the cyan-colored strip in this
  figure. As can be seen, the theoretical region is excluded by some
  experiments (curves labeled 6, 8, 9) but allowed by others (curves
  labeled 1, 2, 3, 5 and parts of curves 4 and 7).

Instead of the Yukawa-type non-Newtonian gravity,
power-law modifications to conventional potential
have also been considered in other contexts \citep{2001PhRvD..64g5010F},
in which case the total potential is written in the form
\begin{equation}
V(r)=-\frac{G_{\infty}m_{1}m_{2}}{r} \left[ 1+\alpha_{N}
  (\frac{r_{0}}{r})^{N-1} \right] .
\label{vr2}
\end{equation}
Here $\alpha_{N}$ is a dimensionless constant and $r_{0}$ corresponds
to a new length scale associated with a non-Newtonian process.  For
example, terms with N=2 and N=3 may be generated by the simultaneous
exchange of two massless scalar particles or two massless pseudoscalar
particles, respectively. (See \cite{2003ARNPS..53...77A} for details.)

Following \cite{2009PhRvD..79l5023K}, the Yukawa-type
non-Newtonian gravity is used in this
paper. \cite{2009PhRvD..79l5023K} suggested that a neutral very
weakly coupled spin-1 gauge U-boson proposed in the super-symmetric
extension of the standard model is a favorite candidate of the
exchanged boson. This light and weakly interacting U-boson has been
used to explain the 511 keV $\gamma$-ray observation from the
galatic bulge
\citep{2003A&A...407L..55J,2004PhRvD..69j1302B,2004PhRvL..92j1301B},
and various experiments in terrestrial laboratories have been
proposed to search for this boson
\citep{2013PhLB..723..388Y}. Since the new bosons contribute to the
EOS of dense matter only through the combination of $g^{2}/\mu^{2}$
\citep{1988IAUS..130..471F}, and the value of $g^{2}/\mu^{2}$ can be
large although both the coupling constant $g$ and the mass $\mu$ of
the light and weakly interacting bosons are
small, the structure of compact
stars may be greatly influenced by the non-Newtonian gravity
effects.

It has been shown by \cite{2009PhRvD..79l5023K} that an
increase of $g$ (a decrease of $\mu$) of scalar bosons has a
negative contribution to pressure, which makes the EOS of
dense matter softer and reduces the maximum mass of a
compact star. By contrast, an increase of $g$ (a decrease of
$\mu$) of vector bosons makes the EOS of dense matter stiffer
and increases the maximum mass of a compact
star. In the following, we will only study the case of vector
bosons since a stiff EOS of SQM is needed for the explanation of
the tidal deformability of GW170817 and the mass of PSR J0740+6620.

The contribution of the
Yukawa correction $V_Y(r)$ of Eq.\ (\ref{vr}) to the energy density
of SQM is obtained by integrating over the quark density distributions
$n_b(\vec x_1)$ and $n_b(\vec x_2)$ contained in a given volume $V$
\citep{2003Natur.421..922L,2009PhRvD..79l5023K,2009PhRvL.103u1102W,2017RAA....17...11L}
\begin{equation}
\epsilon_{Y}=\frac{1}{2V}\int 3n_{b}(\vec{x}_{1})\frac{g^{2}}{4\pi}
\frac{e^{-\mu r}}{r}3n_{b}(\vec{x}_{2})d\vec{x}_{1}d\vec{x}_{2} ,
\label{inte1}
\end{equation}
where $r = |\vec{x}_{1} - \vec{x}_{2}|$. The prefactors of 3 in
front of the quark densities are required since the baryon number of
quarks is $1/3$. Equation (\ref{inte1}) can be evaluated further
since the quark
densities $n_{b}(\vec{x}_{1}) = n_{b}(\vec{x}_{2}) \equiv n_{b}$ are
essentially independent of position
\citep{1986ApJ...310..261A,alcock88,madsen99,2005PrPNP..54..193W}. Moving
$n_{b}$ outside of the integral then leads for the energy of SQM
contained inside of $V=4\pi R^3/3$ (for simplicity taken to be
spherical \footnote{The actual geometry of the volume is
	unimportant since we are only interested in the local
	modification of the energy (Eq.\ (\ref{eq:epsY})) caused by the
	Yukawa term.}) to \citep{2017RAA....17...11L}
\begin{equation}
\epsilon_{Y} = \frac{9}{2}g^{2}n_{b}^{2} \int_{0}^{R} r e^{-\mu r} dr.
\label{inte2}
\end{equation}
Upon carrying out the integration over the spherical volume one
arrives at
\begin{equation}
\epsilon_{Y} = \frac{9}{2} \frac{g^{2}n_{b}^{2}} {\mu^{2}} \left[1-(1+\mu
R) e^{-\mu R} \right] .
\label{ey1}
\end{equation}
Because the system we are considering is in principle very large, we may
take $R\rightarrow \infty$ in Eq.\ (\ref{ey1}) to arrive at
\begin{equation}
\epsilon_{Y} = \frac{9}{2} \frac{g^{2}}{\mu^{2}}n_{b}^{2} .
\label{eq:epsY}
\end{equation}
This analysis shows that the additional contribution to the energy
density from the Yukawa correction, $V_Y$, is simply determined (aside
from some constants) by the number of quarks per volume. The total
energy density of SQM is obtained by adding $\epsilon_Y$ to the
standard expression for the energy density of SQM given by
Eq.\ (\ref{eq:epsQ}), leading to
\begin{equation}
\epsilon = \epsilon_{Q} + \epsilon_{Y} .
\end{equation}
Correspondingly, the extra pressure due to the Yukawa correction is
\begin{equation}
p_{Y}=n_{b}^{2}\frac{d}{dn_{b}}\bigg(\frac{\epsilon_{Y}}{n_{b}}\bigg)
= \frac{9}{2}\frac{g^{2}n_{b}^{2}}{\mu^{2}}
\bigg(1-\frac{2n_{b}}{\mu}\frac{\partial \mu}{\partial n_{b}}\bigg).
\end{equation}
Assuming a constant boson mass independent of the density
\citep{2009PhRvD..79l5023K,2009PhRvL.103u1102W,2017RAA....17...11L},
one obtains
\begin{equation}
p_{Y}=\epsilon_{Y} = \frac{9}{2} \frac{g^{2}}{\mu^{2}}n_{b}^{2}.
\end{equation}
The total pressure including the non-Newtonian gravity (Yukawa) term then reads
\begin{equation}
p = p_{Q} + p_{Y} ,
\end{equation}
where $p_Q$ is given by Eq.\ (\ref{eq:pQ}).

In summary, the full EOS of SQM accounting for the Yukawa correction
is given by $p(\epsilon)$. This quantity constitutes, via the energy-momentum
tensor
\begin{equation}
T^{\alpha\beta} = \left(
\epsilon + p(\epsilon) \right) u^\alpha u^\beta + p(\epsilon) g^{\alpha\beta} ,
\end{equation}
the source term of Einstein's field equation.  The effects of the
Yukawa correction term on compact stellar objects can thus be
explored as usual by solving the Tolman-Oppenheimer-Volkoff (TOV)
equation
\citep{tolman39,oppenheimer39,2009PhRvD..79l5023K,2009PhRvL.103u1102W,2014JPhG...41g5203L}
with $p(\epsilon)$ (the matter equation) serving as an input quantity
\citep{1988IAUS..130..471F}.

\subsection{Strange Quark Star Structure and Tidal Deformability}

In the following, we use geometrized units $G=c=1$ and define the
compactness $\beta$, which is given by $\beta\equiv M/R$.

The dimensionless tidal deformability is defined as $\Lambda\equiv
\lambda/M^{5}$, where $\lambda$ denotes the tidal deformability
parameter, \footnote{The symbol $\lambda$ is also used in
	Eqs.\ (\ref{vr}) and (\ref{mulam}), where it denotes the length scale
	of non-Newtonian gravity. It symbolizes the tidal deformability
	parameter conventionally only in this paragraph.} which can be
expressed in terms of the dimensionless tidal Love number $k_{2}$ as
$\lambda=\frac{2}{3}k_{2}R^{5}$
\citep{2008PhRvD..77b1502F,2008ApJ...677.1216H,2009PhRvD..80h4035D,2010PhRvD..81l3016H}. Thus,
one has
\begin{equation}
\Lambda=\frac{2}{3}k_{2}\beta^{-5}.
\end{equation}

The tidal Love number $k_{2}$ can be calculated using the expression
\citep[e.g.,][]{2016PhR...621..127L,2019JPhG...46c4001W,2020EPJA...56...63W}
\begin{equation}
k_{2}=\frac{8}{5}\frac{\beta^{5}z}{F},
\end{equation}
with
\begin{eqnarray}
z\equiv(1-2\beta)^{2}[2-y_{R}+2\beta(y_{R}-1)]
\label{z}
\end{eqnarray}
and
\begin{eqnarray}
\nonumber
F&\equiv&6\beta(2-y_{R})+6\beta^{2}(5y_{R}-8)+4\beta^{3}(13-11y_{R}) \\
&&  +4\beta^{4}(3y_{R}-2)+8\beta^{5}(1+y_{R})+3z\textrm{ln}(1-2\beta).~~~~
\label{f}
\end{eqnarray}

In Eqs.\ (\ref{z}) and (\ref{f}), $y_{R}\equiv y(R)-4\pi
R^{3}\epsilon_{s}/M$, where $y(R)$ is the value of $y(r)$ at the
surface of the star, and the second term of right hand side exists
because there is a nonzero energy density $\epsilon_{s}$ just inside
the surface of QSs \citep{2010PhRvD..82b4016P}. The quantity $y(r)$ satisfies the differential equation
\begin{equation}
\frac{dy(r)}{dr}=-\frac{y(r)^{2}}{r}-\frac{y(r)-6}{r-2m(r)}-rQ(r),
\label{yr}
\end{equation}
with
\begin{eqnarray}
\nonumber Q(r)&\equiv&4\pi\frac{[5-y(r)]\epsilon(r)+[9+y(r)]p(r)
	+[\epsilon(r)+p(r)]/c_{s}^{2}}{1-2m(r)/r}\\ &&-4\bigg[\frac{m(r)+4\pi
	r^{3}p(r)}{r[r-2m(r)]}\bigg]^{2} ,
\label{Qr}
\end{eqnarray}
where $c_{s}^{2}=dp(r)/d\epsilon(r)$ is the speed of sound. For a
given EOS, Eq.\ (\ref{yr}) can be calculated together with the
TOV equation, i.e.,
\begin{equation}
\frac{dp(r)}{dr}=-\frac{[m(r)+4\pi r^{3}p(r)][\epsilon(r)+p(r)]}{r[r-2m(r)]},
\end{equation}
\begin{equation}
\frac{dm(r)}{dr}=4\pi \epsilon(r) r^{2},
\end{equation}
with the boundary conditions $y(0)$=2, $p(R)$=0, $m(0)$=0 for a given pressure at the center of the star $p(0)$.

\section{Results and Discussions}\label{ResDisc}

\begin{figure}[htb]
	\begin{center}
		\includegraphics[width=0.55\textwidth]{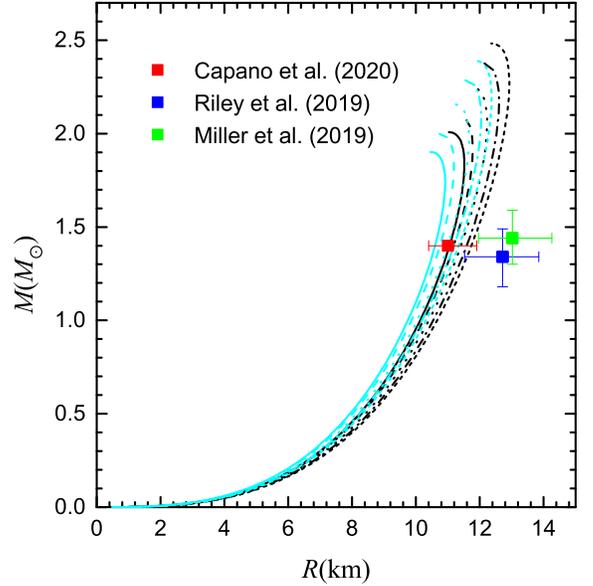}
		\caption{The mass-radius relation of strange
			stars. The black curves are for $B^{1/4}$=142 MeV,
			$\alpha_{S}$=0.2, and the cyan curves are for $B^{1/4}$=146
			MeV, $\alpha_{S}$=0. The solid, dashed, dotted,
			dash-dotted, short-dashed lines are for $g^{2}/\mu^{2}=0$,
			1, 3, 5, and 7 GeV$^{-2}$, respectively. The red data is
			$R_{1.4} = 11.0_{-0.6}^{+0.9}$ km, which is the radius of
			$1.4\, M_{\odot}$ constrained by the observations of GW170817
			\citep{2020NatAs.tmp...42C}. The blue and green data show
			the mass and radius estimates of PSR J0030+0451 derived from
			NICER data by \cite{2019ApJ...887L..21R} ($R =
			12.71_{-1.19}^{+1.14}$ km, $M = 1.34_{-0.16}^{+0.15}\,
			M_{\odot}$) and \cite{2019ApJ...887L..24M} ($R =
			13.02_{-1.06}^{+1.24}$ km, $M =1.44_{-0.14}^{+0.15}\,
			M_{\odot}$).} \label{fig1}
	\end{center}
\end{figure}

\begin{figure*}[tb]
	\begin{center}
		\includegraphics[width=1.0\linewidth]{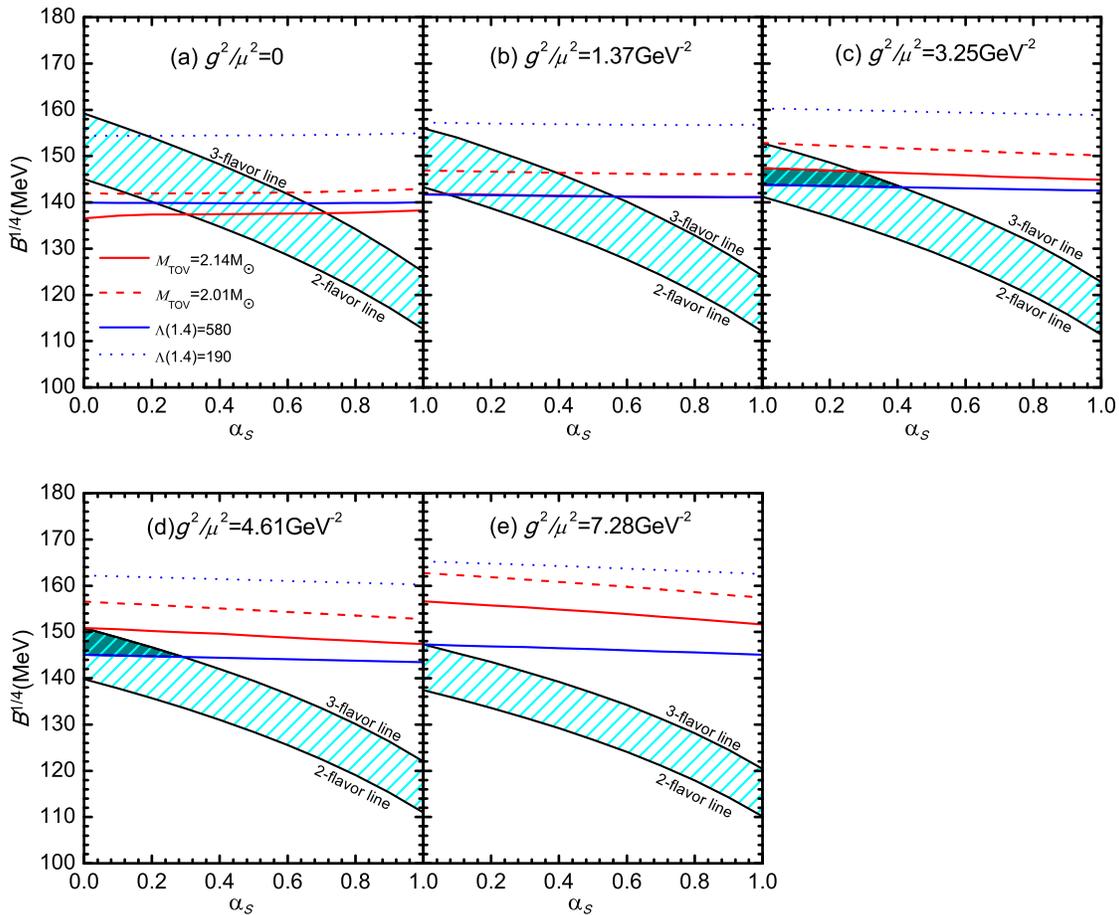}
		\caption{Constraints on $B^{1/4}$ and $\alpha_{S}$ for a strange quark
			mass of $m_{s}=95$ MeV.  The non-Newtonian gravity parameters
			are $g^{2}/\mu^{2}=0$ (a), $g^{2}/\mu^{2}=1.37$ GeV$^{-2}$ (b),
			$g^{2}/\mu^{2}=3.25$ GeV$^{-2}$ (c), $g^{2}/\mu^{2}=4.61$
			GeV$^{-2}$ (d) and $g^{2}/\mu^{2}=7.28$ GeV$^{-2}$ (e).  The dark
			cyan-shadowed regions in panels (c) and (d) indicate the allowed
			parameter spaces.  (See the text for details.)}
		\label{fig2}
	\end{center}
\end{figure*}

By solving the TOV equations, the mass-radius relation of strange
stars for different non-Newtonian gravity parameters is shown in
Fig.\ \ref{fig1}. We choose two sets of parameters, namely,
$B^{1/4}$=142 MeV, $\alpha_{S}$=0.2 and $B^{1/4}$=146 MeV,
$\alpha_{S}$=0, since they can well satisfy both the "2-flavor line"
and "3-flavor line" constraints, which will be shown later in
Fig.\ \ref{fig2}. One can see that the maximum mass of strange stars becomes
larger with the inclusion of the Yukawa (non-Newtonian gravity)
term. We also find that the radius  of a $1.4\, M_{\odot}$ QS is
consistent with the results derived from
GW170817, but PSR J0030+0451 will not be a QS.

\begin{figure*}[tb]
	\begin{center}
		\includegraphics[width=1.0\linewidth]{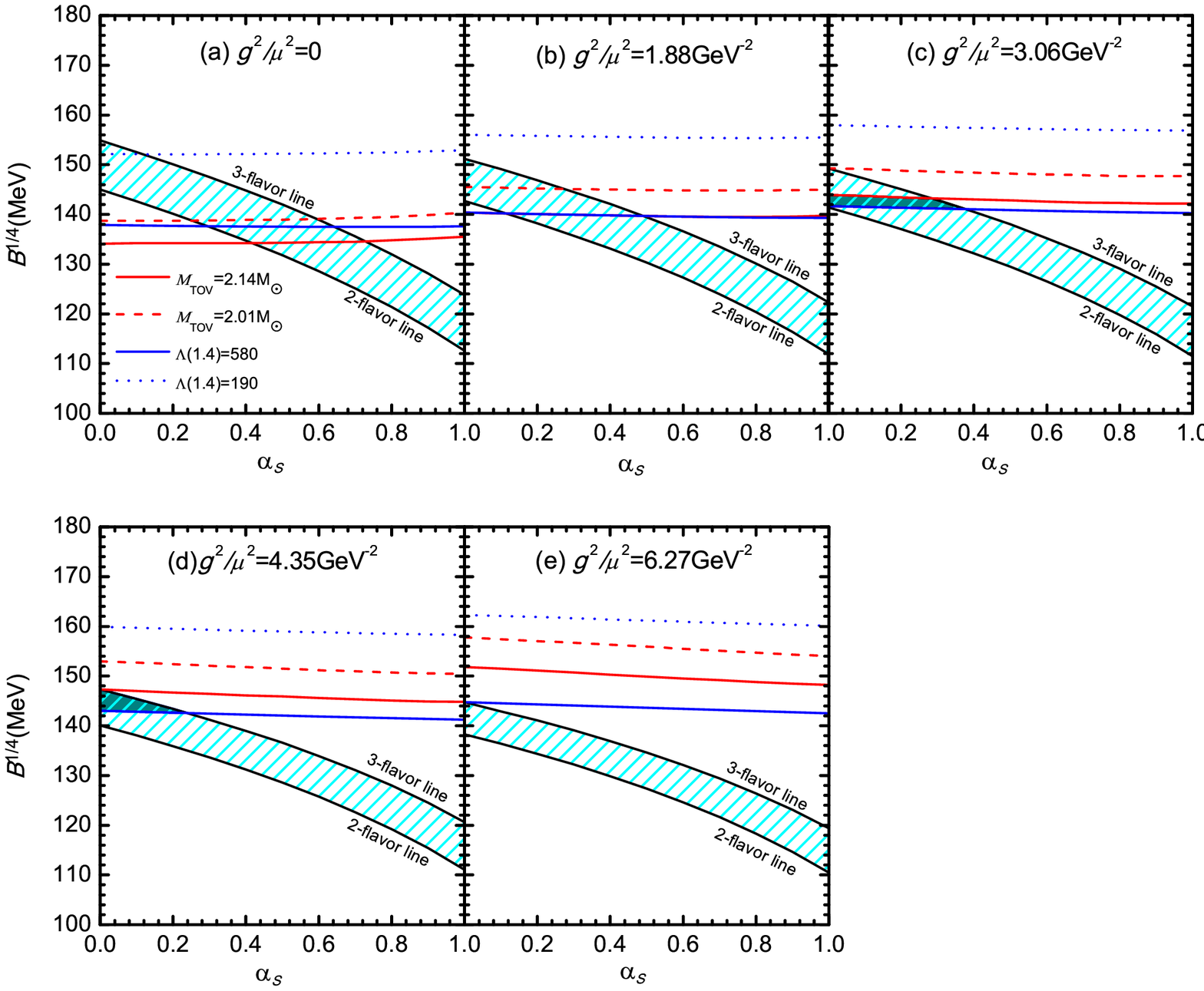}
		\caption{Constraints on $B^{1/4}$ and $\alpha_{S}$ for a strange quark mass of
			$m_{s}=150$ MeV. The non-Newtonian gravity parameters are
			$g^{2}/\mu^{2}=0$ (a), $g^{2}/\mu^{2}=1.88$ GeV$^{-2}$ (b),
			$g^{2}/\mu^{2}=3.06$ GeV$^{-2}$ (c), $g^{2}/\mu^{2}=4.35$
			GeV$^{-2}$ (d), and $g^{2}/\mu^{2}=6.27$ GeV$^{-2}$ (e).}
		\label{fig3}
	\end{center}
\end{figure*}

We calculate the allowed parameter space of the SQM model
according to the following constraints
\citep[e.g.,][]{1997JPhG...23.2029S,2011ApJ...740L..14W,2015RAA....15..871P,2018PhRvD..97h3015Z}:
First, the existence of QSs is based on the idea that the presence of
strange quarks lowers the energy per baryon of
a mixture of $u$, $d$ and $s$ quarks in beta equilibrium
below the energy of the most stable atomic nucleus, $^{56}$Fe
($E/A\sim 930$ MeV) \citep{1984PhRvD..30..272W}. This constraint
results in the 3-flavor line shown in Figs.\ \ref{fig2} and
\ref{fig3}.

The second constraint is given by assuming that non-strange quark
matter (i.e., two-flavor quark matter made of only $u$ and $d$ quarks)
in bulk has an energy per baryon higher than the one of $^{56}$Fe,
plus a 4 MeV correction coming from surface effects
  \citep{1984PhRvD..30.2379F, madsen99, 2011ApJ...740L..14W,
    2018PhRvD..97h3015Z}. By imposing $E/A\geq 934$ MeV on
non-strange quark matter, one ensures that atomic nuclei do not
dissolve into their constituent quarks. This leads to the 2-flavor
lines in Figs.\ \ref{fig2} and \ref{fig3}.  The shaded areas between
the 3-flavor lines and the 2-flavor lines in Figs.\ \ref{fig2} and
\ref{fig3} show the (allowed) $B^{1/4}$--$\alpha_s$ parameter regions
where both constraints on the energy per baryon described just above
are fulfilled.

The third constraint is that the maximum mass of QSs
must be greater than the mass of PSR J0740+6620, $M_{\rm max} \geq
2.14\, M_{\odot}$. By employing this constraint, the allowed
parameter space is limited to the region below the red solid lines
in Figs.\ \ref{fig2} and \ref{fig3}. The red dashed lines in
Figs.\ \ref{fig2} and \ref{fig3} correspond to a stellar mass of
$2.01\, M_{\odot}$ and are shown for comparison.

The last constraint follows from $\Lambda(1.4)\leq 580$, where
$\Lambda(1.4)$ is the dimensionless tidal deformability of a
$1.4\, M_{\odot}$ star. The parameter space satisfies this constraint
corresponds to the region above the blue solid lines in
Figs.\ \ref{fig2} and \ref{fig3}. The blue dotted lines shown in
these figures correspond to a tidal deformability of $\Lambda(1.4)=190$.

By imposing all four constraints mentioned above, the allowed
$B^{1/4}$--$\alpha_s$ parameter space of the SQM model
considered in this paper is restricted to the dark cyan-shadowed
regions shown in Figs.\ \ref{fig2}(c), \ref{fig2}(d),
\ref{fig3}(c), and \ref{fig3}(d), which are obtained for
non-Newtonian gravity parameter values of $g^{2}/\mu^{2}=3.25$
GeV$^{-2}$, $g^{2}/\mu^{2}=4.61$ GeV$^{-2}$, $g^{2}/\mu^{2}=3.06$
GeV$^{-2}$, and $g^{2}/\mu^{2}=4.35$ GeV$^{-2}$,
respectively. For all other cases studied in our paper, an
overlapping region where all four constraints are simultaneously
satisfied does not exist. This is selectively shown in
Figs.\ \ref{fig2}(a), \ref{fig2}(b), and \ref{fig2}(e) which
correspond to $g^{2}/\mu^{2}=0$, $g^{2}/\mu^{2}=1.37$ GeV$^{-2}$,
and $g^{2}/\mu^{2}=7.28$ GeV$^{-2}$, respectively.  Figures
\ref{fig3}(a), \ref{fig3}(b), and \ref{fig3}(e) illustrate the
situation for a strange quark mass of 150 MeV. Here, the
corresponding non-Newtonian gravity parameter values are
$g^{2}/\mu^{2}=0$, $g^{2}/\mu^{2}=1.88$ GeV$^{-2}$, and
$g^{2}/\mu^{2}=6.27$ GeV$^{-2}$, respectively.

From Figs.\ \ref{fig2}(a) and \ref{fig3}(a), one sees that for the case
of $g^{2}/\mu^{2}=0$, the constraints $M_{\rm max}\geq 2.14\, M_{\odot}$
and $\Lambda(1.4)\leq 580$ can not be be satisfied simultaneously.
This situation continues as the value of $g^{2}/\mu^{2}$ becomes
bigger until it is as large as 1.37 GeV$^{-2}$ for a strange quark mass of $m_{s}=95$ MeV
(1.88 GeV$^{-2}$ for $m_{s}=150$ MeV), in which case the
$M_{\rm max}=2.14\, M_{\odot}$ lines almost completely coincide with the
$\Lambda(1.4)=580$ lines (see Figs.\ \ref{fig2}(b) and \ref{fig3}(b)).

Depending on the value of the strange quark mass, the allowed
parameter space vanished entirely for $g^{2}/\mu^{2} > 7.28$
GeV$^{-2}$ or $g^{2}/\mu^{2} > 6.27$ GeV$^{-2}$, as shown in
Figs.\ \ref{fig2}(e) and \ref{fig3}(e).

It is necessary to focus on Figs.\ \ref{fig2}(b) and \ref{fig2}(d)
(Figs.\ \ref{fig3}(b) and \ref{fig3}(d)) once again. In
Fig.\ \ref{fig2}(b) (Fig.\ \ref{fig3}(b)), the $M_{max}=2.14\,
M_{\odot}$ line and the $\Lambda(1.4)=580$ line almost completely
overlap. These two lines cut across the "3-flavor line" at the point
where the bag constant has a value of $B^{1/4}=141.3$ MeV
($B^{1/4}=139.7$ MeV) and the strong coupling constant
has a value of $\alpha_{S}=0.56$ ($\alpha_{S}=0.49$).
Therefore, one can draw the conclusion that by
considering the four constraints discussed
in this paper, the lower limit of $B^{1/4}$ is 141.3 MeV (139.7 MeV)
, and the upper limit of $\alpha_{S}$ is 0.56 (0.49)
. On the other hand, in Fig.\ \ref{fig2}(d)
(Fig.\ \ref{fig3}(d)), the $M_{max}= 2.14\, M_{\odot}$ line meets
"3-flavor line" and the longitudinal coordinate-axis at
$B^{1/4} =150.9$ MeV ($B^{1/4}=147.3$ MeV), which
suggests that the upper limit of $B^{1/4}$ is 150.9 MeV (147.3 MeV).

\section{Summary}\label{Summary}

In this paper, we have investigated the effects of non-Newtonian
gravity on the properties of QSs and constrained the parameter
space of the SQM model using observations related to PSR
J0740+6620 and GW170817. It is found that these observations can not
be explained by the SQM model studied in this paper if
non-Newtonian gravity effects are not included. In other words, the
existence of QSs would be ruled out in this case. Considering the
non-Newtonian gravity effects, regions in the $B^{1/4}$--$\alpha_{S}$
parameter space have been established for which SQM exists (i.e., is
absolutely stable) and the EOS associated with such matter leads to
properties of compact stars that are in agreement with
observation. The constraints on the bag constant and the strong coupling constant
depend on the mass of the strange quark. For a strange quark mass of $m_{s}=95$ MeV,
$141.3~ {\rm MeV} \leq B^{1/4} \leq 150.9~ {\rm MeV}$ and $\alpha_{S}\leq 0.56$.
While, for a strange quark mass of $m_{s}=150$ MeV,
139.7 MeV$\leq B^{1/4}\leq$ 147.3 MeV and $\alpha_{S}\leq 0.49$.

  Moreover, as shown in Fig.\ \ref{fig0}, the ranges of the
  non-Newtonian gravity parameter (1.37 GeV$^{-2}\leq g^{2}/\mu^{2}\leq$ 7.28
  GeV$^{-2}$ for $m_{s}=95$ MeV, and 1.88 GeV$^{-2}\leq
  g^{2}/\mu^{2}\leq$ 6.27 GeV$^{-2}$ for $m_{s}=150$ MeV) agree well
  with  the constraints set by some, but not all, experiments.

As shown in this paper, the possible existence of (absolutely
stable) QSs is impacted by non-Newtonian gravity.  The existence of
QSs, therefore, may constrain the non-Newtonian gravity parameter
$g^{2}/\mu^{2}$. And for a given value of $g^{2}/\mu^{2}$, the
allowed parameter space of SQM ($B^{1/4}$ and $\alpha_{S}$) can be
fixed. Our results are intimately related to the maximum mass of neutron stars,
their radii, and the upper
limit of the tidal deformability of neutron stars. In light of
the rapidly increasing data on the properties of such objects
provided by NICER and gravitational-wave interferometers, this
connection should definitely be explored further.  The conclusions
drawn in this investigation are based on the standard MIT bag model
of SQM. It would be interesting to carry out a similar investigation
which is based on other SQM models such as the quasi-particle model (for a brief
review, see \cite{2015PhRvD..92b5025X}). Finally, we mention that
the existence of QSs can (and should) also be investigated in the
framework of other, alternative theories of gravity
\citep[e.g.,][]{2019EPJA...55..117L}.

\acknowledgements We are especially indebted to the
  anonymous referee for his/her useful suggestions that helped us to
  improve the paper. This work is supported by the Fundamental
Research Funds for the Central Universities (Grant No. CCNU19QN063)
and the Scientific Research Program of the National Natural Science
Foundation of China (NSFC, Grant Nos.\ 11447012 and
11773011). F.\ W.\ is supported through the U.S.\ National Science
Foundation under Grants PHY-1714068 and PHY-2012152.


\begin{thebibliography}{}
	
	\bibitem[Abbott et al.(2017a)]{2017PhRvL.119p1101A}Abbott, B. P., et
	al.\ 2017a, Phys. Rev. Lett., 119, 161101
	
	\bibitem[Abbott et al.(2017b)]{2017ApJ...848L..12A}Abbott, B. P., et
	al.\ 2017b, \apjl, 848, L12
	
	\bibitem[Abbott et al.(2018)]{2018PhRvL.121p1101A}Abbott, B. P., et
	al.\ 2018, Phys. Rev. Lett., 121, 161101
	
	\bibitem[Adelberger et al.(2003)]{2003ARNPS..53...77A}Adelberger,
	E. G., Heckel, B. R., \& Nelson, A. E.\ 2003,
	Ann. Rev. Nucl. Part. Sci., 53, 77
	
	\bibitem[Adelberger et al.(2009)]{2009PrPNP..62..102A}Adelberger,
	E. G., Gundlach, J. H., Heckel, B. R., Hoedl, S., \& Schlamminger,
	S.\ 2009, Prog. Part. Nucl. Phys., 62, 102
	
	\bibitem[Alcock et al.(1986)]{1986ApJ...310..261A}Alcock, C., Farhi,
	E., \& Olinto, A.\ 1986, \apj, 310, 261
	
	\bibitem[Alcock et al.(1988)]{alcock88}Alcock, C., \& Olinto, A. V.\ 1988,
	Ann. Rev. Nucl. Part. Sci., 38, 161
	
	\bibitem[Antoniadis(2013)]{2013Sci...340..448A}Antoniadis, J., et
	al.\ 2013, Science, 340, 1233232
	
	\bibitem[Baiotti(2019)]{2019PrPNP.10903714B}Baiotti, L.\ 2019,
	Prog. Part. Nucl. Phys., 109, 103714
	
	\bibitem[Boehm et al.(2004a)]{2004PhRvD..69j1302B} Boehm, C., Fayet,
	P., \& Silk, J.\ 2004a, Phys. Rev. D, 69, 101302
	
	\bibitem[Boehm et al.(2004b)]{2004PhRvL..92j1301B} Boehm, C., Hooper,
	D., Silk, J., Casse, M., \& Paul, J.\ 2004b, Phys. Rev. Lett., 92,
	101301
	
	\bibitem[Capano et al.(2020)]{2020NatAs.tmp...42C}Capano, C. D., et
	al.\ 2020, Nat. Astron., 4, 625
	
	\bibitem[Chen et al.(2016)]{Chen2016} 
	Chen,Y.-J., Tham, W. K., Krause, D. E., López, D., Fischbach, E., \& Decca, R. S.\ 2016, Phys. Rev. Lett., 116, 221102
	
	\bibitem[Cromartie et al.(2020)]{2020NatAs...4...72C}Cromartie, H. T.,
	et al.\ 2020, Nat. Astron., 4, 72
	
	\bibitem[Damour \& Nagar(2009)]{2009PhRvD..80h4035D}Damour, T., \&
	Nagar, A.\ 2009, Phys. Rev. D, 80, 084035
	
	\bibitem[Farhi \& Jaffe(1984)]{1984PhRvD..30.2379F}Farhi, E., \&
	Jaffe, R. L.\ 1984, Phys. Rev. D, 30, 2379
	
	\bibitem[Fayet(1980)]{1980PhLB...95..285F}Fayet, P.\ 1980,
	Phys. Lett. B, 95, 285
	
	\bibitem[Fayet(1981)]{1981NuPhB.187..184F}Fayet, P.\ 1981,
	Nucl. Phys. B, 187, 184
	
	\bibitem[Fischbach et al.(2001)]{2001PhRvD..64g5010F}Fischbach, E.,
	Krause, D. E., Mostepanenko, V. M., \& Novello, M.\ 2001, Phys. Rev. D, 64, 075010
	
	\bibitem[Fischbach \& Talmadge(1999)]{1999snng.book.....F}Fischbach,
	E., \& Talmadge, C. L.\ 1999, \emph{The Search for Non-Newtonian
		Gravity} (Springer-Verlag, Inc., New York)
	
	\bibitem[Flanagan \& Hinderer(2008)]{2008PhRvD..77b1502F}Flanagan,
	E. E., \& Hinderer, T.\ 2008, Phys. Rev. D, 77, 021502(R)
	
	\bibitem[Fujii(1971)]{1971NPhS..234....5F}Fujii, Y.\ 1971, Nature
	Physical Science, 234, 5
	
	\bibitem[Fujii(1988)]{1988IAUS..130..471F}Fujii, Y.\ 1988, \emph{Large
		Scale Structures of the Universe}, edited by J. Audouze et
	al. (International Astronomical Union, Dordrecht: Kluwer), p.471
	
	\bibitem[Haensel et al.(1986)]{1986A&A...160..121H}Haensel, P.,
	Zdunik, J. L., \& Schaefer, R.\ 1986, \aap, 160, 121
	
	\bibitem[Hinderer(2008)]{2008ApJ...677.1216H}Hinderer, T.\ 2008, \apj,
	677, 1216
	
	\bibitem[Hinderer et al.(2010)]{2010PhRvD..81l3016H}Hinderer, T.,
	Lackey, B. D., Lang, R. N., \& Read, J. S.\ 2010, Phys. Rev. D, 81,
	123016
	
	\bibitem[Jean et al.(2003)]{2003A&A...407L..55J} Jean, P., et
	al. 2003, \aap, 407, L55
	
	\bibitem[Kamiya et al.(2015)]{Kamiya2015}Kamiya, Y., Itagami, K., Tani, M., Kim, G. N., \& Komamiya, S.\ 2015, Phys. Rev. Lett., 114, 161101
		
	\bibitem[Kamyshkov et al.(2008)]{Kamyshkov2008}Kamyshkov, Y., Tithof, J., \& Vysotsky, M.\ 2008, Phys. Rev. D, 78, 114029
	
	\bibitem[Klimchitskaya et al.(2020)]{Klimchitskaya2020} Klimchitskaya, G. L., Kuusk, P., \& Mostepanenko, V. M.\ 2020, Phys. Rev. D, 101, 056013
	
	\bibitem[Krivoruchenko et
	al.(2009)]{2009PhRvD..79l5023K}Krivoruchenko, M. I., \v{S}imkovic,
	F., \& Faessler, A.\ 2009, Phys. Rev. D, 79, 125023
	
	\bibitem[Lai et al.(2019)]{2019EPJA...55...60L}Lai, X., Zhou, E., \&
	Xu, R.\ 2019, Eur. Phys. J. A, 55, 60
	
	\bibitem[Lai et al.(2018)]{2018RAA....18...24L}Lai, X.-Y., Yu,
	Y.-W., Zhou, E.-P., Li, Y.-Y., \& Xu, R.-X.\ 2018,
	Res. Astron. Astrophys., 18, 24
	
	\bibitem[Lattimer \& Prakash(2016)]{2016PhR...621..127L}Lattimer,
	J. M., \& Prakash, M.\ 2016, Phys. Rep., 621, 127
	
	\bibitem[Li et al.(2019)]{2019EPJA...55..117L}Li, B.-A., Krastev,
	P. G., Wen, D.-H., \& Zhang, N.-B.\ 2019, Eur. Phys. J. A, 55, 117
	
	\bibitem[Lin et al.(2014)]{2014JPhG...41g5203L}Lin, W., Li, B.-A.,
	Chen, L.-W., Wen, D.-H., \& Xu, J.\ 2014, J. Phys. G:
	Nucl. Part. Phys., 41, 075203
	
	\bibitem[Long et al.(2003)]{2003Natur.421..922L}Long, J. C., et
	al.\ 2003, Nature, 421, 922
	
	\bibitem[Lu et al.(2017)]{2017RAA....17...11L}Lu, Z.-Y., Peng,
	G.-X., \& Zhou, K.\ 2017, Res. Astron. Astrophys., 17, 11
	
	\bibitem[Madsen(1999)]{madsen99}Madsen, J. 1999, Lecture Notes in
	Physics, 516, 162
	
	\bibitem[Miller et al.(2019)]{2019ApJ...887L..24M}Miller, M. C., et
	al.\ 2019, \apj, 887, L24
	
	\bibitem[Murata \& Tanaka(2015)]{2015CQGra..32c3001M}Murata, J., \&
	Tanaka, S.\ 2015, Classical Quantum Gravity, 32, 033001
	
	\bibitem[Olive et al.(2014)]{2014ChPhC..38i0001O}Olive, K. A., et
	al. (Particle Data Group)\ 2014, Chin. Phys. C, 38, 090001
	
	\bibitem[Oppenheimer \& Volkoff(1939)]{oppenheimer39}Oppenheimer, J. R., \&
	Volkoff, G. M.\ 1939, Phys. Rev., 55, 374
	
	\bibitem[Orsaria et al.(2019)]{2019JPhG...46g3002O}Orsaria, M. G, et
	al.\ 2019, J. Phys. G: Nucl. Part. Phys., 46, 073002
	
	\bibitem[Pi et al.(2015)]{2015RAA....15..871P}Pi, C.-M., Yang,
	S.-H., \& Zheng, X.-P. \ 2015, Res. Astron. Astrophys., 15, 871
	
	\bibitem[Pokotilovski(2006)]{Pokotilovski2006}Pokotilovski, Y. N.\ 2006, Phys. At. Nucl., 69, 924
		
	\bibitem[Postnikov et al.(2010)]{2010PhRvD..82b4016P}Postnikov, S.,
	Prakash, M., \& Lattimer, J. M.\ 2010, Phys. Rev. D, 82, 024016
	
	\bibitem[Raithel(2019)]{2019EPJA...55...80R}Raithel, C. A.\ 2019,
	Eur. Phys. J. A, 55, 80
	
	\bibitem[Riley et al.(2019)]{2019ApJ...887L..21R}Riley, T. E., et
	al.\ 2019, \apj, 887, L21
	
	\bibitem[Schaab et al.(1997)]{1997JPhG...23.2029S}Schaab, C., Hermann,
	B., Weber, F., \& Weigel, M. K.\ 1997, J. Phys. G:
	Nucl. Part. Phys., 23, 2029
	
	\bibitem[Sulaksono et al.(2011)]{2011MPLA...26..367S}Sulaksono, A.,
	Marliana, \& Kasmudin\ 2011, Mod. Phys. Lett. A, 26, 367
	
	\bibitem[Tanabashi et al.(2018)]{Tanabashi18}Tanabashi, M. et al.\ 2018,
	Phys. Rev. D, 98, 030001
	
	\bibitem[Tolman(1939)]{tolman39} Tolman, R. C.\ 1939, Phys. Rev., 55,
	364
	
	\bibitem[Weber(2005)]{2005PrPNP..54..193W}Weber, F. 2005,
	Prog. Part. Nucl. Phys., 54, 193
	
	\bibitem[Wei et al.(2019)]{2019JPhG...46c4001W}Wei, J. B., Figura, A.,
	Burgio, G. F., Chen, H., \& Schulze, H.-J.\ 2019, J. Phys. G:
	Nucl. Part. Phys., 46, 034001
	
	\bibitem[Wei et al.(2020)]{2020EPJA...56...63W}Wei, J. B., Lu, J. J.,
	Burgio, G. F., Li, Z. H., \& Schulze, H.-J.\ 2020, Eur. Phys. J. A,
	56, 63
	
	\bibitem[Weissenborn et al.(2011)]{2011ApJ...740L..14W}Weissenborn,
	S., Sagert, I., Pagliara, G., Hempel, M., \& Schaffner-Bielich,
	J.\ 2011, \apj, 740, L14
	
	\bibitem[Wen et al.(2009)]{2009PhRvL.103u1102W}Wen, D.-H., Li,
	B.-A., \& Chen, L.-W.\ 2009, Phys. Rev. Lett., 103, 211102
	
	\bibitem[Witten(1984)]{1984PhRvD..30..272W}Witten, E.\ 1984,
	Phys. Rev. D, 30, 272
	
	\bibitem[Xu et al.(2013)]{Xu2013}Xu, J., Li, B.-A., Chen, L.-W., \& Zheng, H.\ 2013, J. Phys. G:
	Nucl. Part. Phys., 40, 035107
	
	\bibitem[Xu et al.(2015)]{2015PhRvD..92b5025X}Xu, J. F., Peng, G.-X.,
	Liu, F., Hou, D.-F., \& Chen, L.-W.\ 2015, Phys. Rev. D, 92,
	025025
	
	\bibitem[Yong et al.(2013)]{2013PhLB..723..388Y} Yong, G.-C., \& Li,
	B.-A.\ 2013, Phys. Lett. B, 723, 388
	
	\bibitem[Zhou et al.(2018)]{2018PhRvD..97h3015Z}Zhou, E.-P., Zhou,
	X., \& Li, A.\ 2018, Phys. Rev. D, 97, 083015
	

\end{thebibliography}
\end{document}